\documentclass[3p,onecolumn,times]{elsarticle}

\usepackage[hidelinks]{hyperref}

\usepackage{graphicx}
\usepackage{amsmath}   

\begin{document}

\begin{frontmatter}

\title{ The Fluorescence Camera of the POEMMA-Balloon with Radio (PBR): Design and Scientific goals }

\author[1]{Matteo Battisti\corref{cor1} }%
\ead{battisti@apc.in2p3.fr}

\author[2]{Johannes Eser}

\author[2]{George Filippatos}

\author[3]{Angela Olinto}

\author[4]{Giuseppe Osteria}

\author[1]{Etienne Parizot}

\author{for the JEM-EUSO Collaboration}

 \cortext[cor1]{Corresponding author}

\affiliation[1]{organization={ Université Paris Cité, CNRS, Astroparticule et Cosmologie, F-75013 Paris, France}} 
\affiliation[2]{organization={University of Chicago},  
                 city={Illinois}, 
                 country={U.S.A.}}
\affiliation[3]{organization={Columbia  University},  
                 city={New York}, 
                 country={U.S.A}}
                 
\affiliation[4]{organization={INFN Napoli}, 
                 country={Italy}}

\begin{abstract}
The POEMMA-Balloon with Radio (PBR) is a proposed payload to fly on a NASA Super Pressure Balloon (SPB). It will act as a pathfinder of the Probe Of Extreme Multi-Messenger Astrophysics (POEMMA) detector.
PBR will consist of an innovative hybrid focal surface featuring a Fluorescence Camera (FC, based on Multi-Anode Photomultiplier Tubes [MAPMTs], 1.05~$\mu$s time resolution) and a Cherenkov Camera (based on SiPMs, 10~ns time resolution), both mounted on the same tiltable frame that can point from nadir up to 13$^\circ$ above the horizon.
The FC's main scientific goal is to observe, for the first time, the fluorescence emission of Extensive Air Showers produced by Ultra-High Energy Cosmic Rays from sub-orbital altitudes. This measurement will validate the detection strategy for future space-based missions, such as POEMMA. As a secondary goal, the FC will perform a search for macroscopic dark matter through slowly evolving showers that will leave a signal similar to (but distinct from) a meteor.
PBR targets a launch in 2027 as a payload of an ultra-long duration balloon flight with a duration of up to 100 days.
\end{abstract}

\end{frontmatter}


\section{The Fluorescence Camera (FC)}
\label{sec:hardware}

\begin{figure}
    \centering
    \includegraphics[width=.8\textwidth]{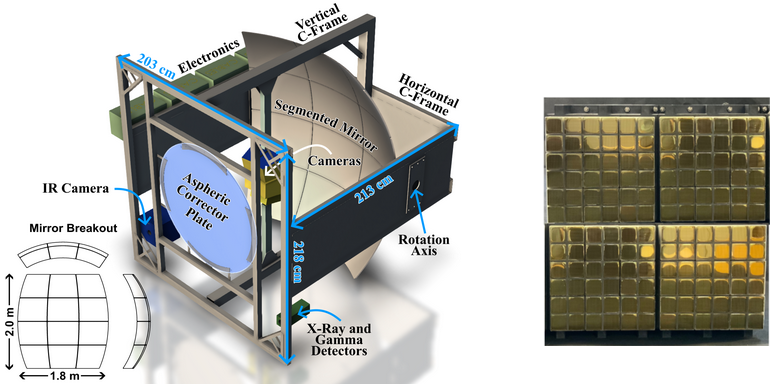}
    \caption{\textbf{Left:} Design of the tiltable frame that hosts the optical system, the FC and CC focal planes and the ancillary devices (infrared cameras, X and gamma ray detectors). \textbf{Right:} 4 PDMs from EUSO-SPB2 arranged in a 2$\times$2 configurations, as intended in PBR.}
    \label{fig:PBR_structure_and_mirrors}
\end{figure}

The PBR FC design is based on the technology developed over the last decade within the JEM-EUSO (Joint Exploratory Missions for Extreme Universe Space Observatory) collaboration and,
in particular, on the design of the SPB2 Fluorescence Telescope (FT) \cite{SPB2-FT}. The optical system consists of a 1.1~m aperture\footnote{This represent an increase of 21\% in the aperture area with respect to the EUSO-SPB2 FT. Once the larger obstruction of the cameras is taken into account, the net gain on the collection area is $\sim$16\%.} Schmidt telescope with 12 mirror segments in 3$\times$4 configuration (Fig.~\ref{fig:PBR_structure_and_mirrors}, left).
The focal plane will be made of 4 Photo Detection Modules (PDMs) arranged in a 2$\times$2 configuration\footnote{This results in an increase of the total field of view with respect to the EUSO-SPB2 FT, whose focal plane consisted of 3 PDMs in a single line.} (Fig.~\ref{fig:PBR_structure_and_mirrors}, right). A PDM is the base of the cameras of the JEM-EUSO detectors, consisting of a 6$\times$6 array of 64-channel MAPMTs (Hamamatsu R11265), for a total of 2304 pixels per PDM. 
The PDMs are based upon stand-alone elements called Elementary Cells (ECs) integrating in a single package 4 MAPMTs, the High Voltage provider (based on Cockcroft–Walton circuit delivering up to 1100~V) and 4 custom-designed SPACIROC-3 ASICs \cite{SPACIROC3}. The custom ASIC performs single photo electron counting on each pixel as well as charge integration on groups of 8 pixels to measure extremely bright and/or fast signals. The two different data acquisition modes will run in parallel and will have independent dedicated trigger logics, as it was in the previous balloon iteration EUSO-SPB2 \cite{SPB2} \cite{SPB2_trigger}.
The time resolution of the FC (both modes) will be of 1.05 $\mu$s (referred as 1 Gate Time Unit, or GTU), with a double pulse resolution (defined as the minimum time distance required between two photo-electrons to be counted as two different photons) of $\sim$5-10~ns. The total field of view will be 24$^\circ \times 24^\circ$, 
for a pixel size on ground of 115~ m.

\section{The calibration procedure}
\label{sec:calibration}
Prior to the launch, the focal surface is calibrated at MAPMT, EC and PDM level. Several features are measured for each pixel, including efficiency, gain, physical size, uniformity, wavelength and
High-Voltage dependence, double pulse resolution, and crosstalk (which results to be negligible).
\begin{figure}[h]
    \centering
    \includegraphics[width=.8\textwidth]{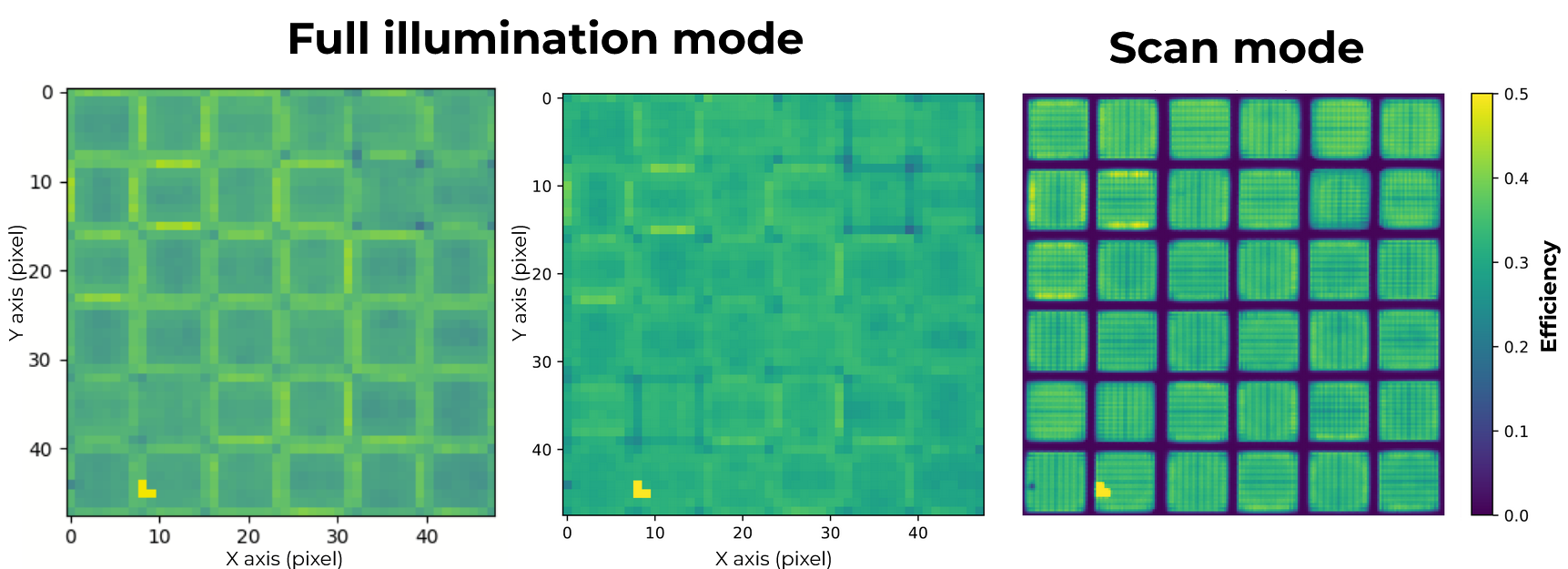}
    \caption{\textbf{Left:} (Apparent) Efficiency map of an EUSO-SPB2 PDM obtained in full illumination mode. On the left (on the right) without (with) compensation for the different physical sizes of the pixels. \textbf{Right:} Efficiency obtained through a high precision scan of the same PDM. }
    \label{fig:calibration_at_APC}
\end{figure}
The calibration is performed in two distinct ways. In full illumination mode (Fig.~\ref{fig:calibration_at_APC}, left) the entire PDM is lighted with a uniform and controlled light coming from an integrating sphere. The apparent\footnote{It is called ``Apparent" efficiency because it is proportional not only to the intrinsic pixel efficiency, but also to the physical size of the pixel. As a result, larger pixels appears to have higher efficiency.} efficiency is obtained as the ratio between the counts measured by each pixel and the photons received by the photocathode. In scan mode, instead, a calibrated light source is installed in front of the camera and moved by a motor with a sub-pixel precision. The obtained map (Fig.~\ref{fig:calibration_at_APC}, right) highlights details of the internal structure of the MAPMTs, the gap between adjacent MAPMTs and shows the absolute efficiency of each part of the PDM.
The exact same procedure has been applied to the EUSO-SPB2 PDMs that showed a remarkable uniformity in terms of efficiency and a high enough gain to be used in photon counting technique. 

\section{The Expected results in UHECR detection}
\label{sec:expected_results}
\begin{figure}[h]
    \centering
    \includegraphics[width=.8\textwidth]{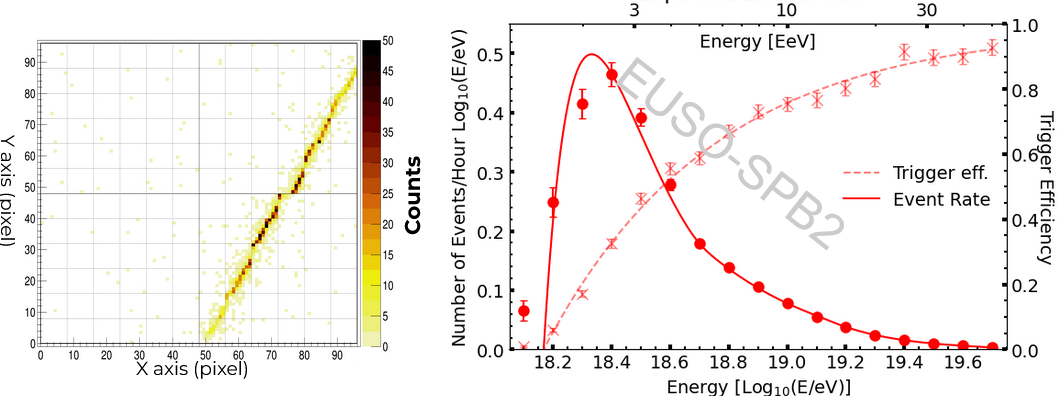}
    \caption{\textbf{Left:} Simulation of a $2\times10^{19}$ UHECR observed by PBR FC. The color scale represents the sum of the counts over 128 GTUs. \textbf{Right:} Trigger efficiency and rate of UHECRs expected for the EUSO-SPB2 Fluorescence Telescope as a function of energy, based on simulation results.}
\label{fig:PBR_event_and_SPB2_trigger_efficiency}
\end{figure}

The main goal of the FC is the observation from sub-orbital altitudes, for the first time, of the fluorescence emission of Extensive Air Showers produced by Ultra-High Energy Cosmic Rays (UHECRs). Thanks to its time resolution, PBR will observe UHECRs as bright spots propagating in a straight line, from which it will be possible to obtain information on the incoming direction and the energy of the primary particle (Fig.~\ref{fig:PBR_event_and_SPB2_trigger_efficiency}, left).  
The previous balloon iteration, EUSO-SPB2, was expected to observe 0.12 events/live hour above $1.5\times10^{18}$~eV ($\sim$1 event every 2 nights), 10\% of which reconstructable (geometry, energy, X$_\text{max}$\footnote{X$_\text{max}$ is defined as the atmospheric depth where the energy deposit profile of the secondary particles reaches its maximum.}) (Fig.~\ref{fig:PBR_event_and_SPB2_trigger_efficiency}, right) \cite{SPB2_performance}.
Preliminary estimations indicate that PBR will detect 50\% more events per live hour due to a lower energy threshold (given by the larger entrance pupil) and a larger field of view (4 PDMs instead of 3). Moreover, when the Cherenkov Camera points above or below the Earth's limb, the FC will keep working in tilted mode, greatly increasing the exposure for UHECRs at the highest energies, at the cost of an increase of the energy threshold. The same observation strategy is foreseen for POEMMA \cite{POEMMA}.

\section{The search for Macroscopic dark matter (MACROs)}
\label{sec:MACROs}
\begin{figure}[h]
    \centering
    \includegraphics[width=.8\textwidth]{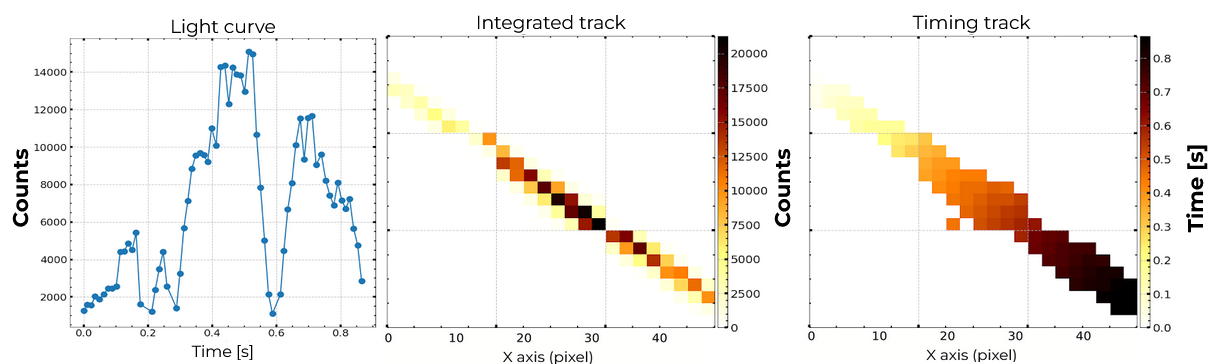}
    \caption{A meteor observed by EUSO-SPB2 during the field test. \textbf{Left:} Evolution of the counts over the PDM as a function of time. \textbf{Center:} Track produced by the meteor. The color scale represent the sum of the counts over 128 GTUs. \textbf{Right:} Time evolution of the signal. The color scale represent the time in which a pixel observes the signal. The signal lasts for almost 1~s. }
    \label{fig:SPB2_meteor}
\end{figure}

The term ``Macroscopic dark matter" (MACROs) includes a large zoology of phenomena that could explain the dark matter mystery with the existence of very massive and very rare (namely with low density in the universe) elements such as nuclearites, strange quark nuggets or primordial black holes. Those phenomena should move at speeds well below the speed of light, depositing significant amounts of energy in the atmosphere similarly to meteors. PBR can look for such an experimental signature, which is similar but clearly distinct from the trace of a meteor\footnote{Moreover, the vast majority of meteors burn in the atmosphere above 33~km, the flight altitude of Super Pressure Balloons (SPBs).}, and will set limits on slow-moving objects and MACROs candidates. The ability of the FC to detect slow-moving object is confirmed by the systematic observation of meteors performed by Mini-EUSO from space \cite{Mini-EUSO_meteors} and by the meteors detected during the EUSO-SPB2 field test from ground (Fig.~\ref{fig:SPB2_meteor}).

\section*{Acknowledgements}

The authors would like to acknowledge the support by NASA award 80NSSC22K1488, by the French space agency CNES and the Italian Space agency ASI. We also acknowledge the invaluable contributions of the administrative and technical staffs at our home institutions.
This research used resources of the National Energy Research Scientific Computing Center (NERSC), a U.S. Department of Energy Office of Science User Facility operated under Contract No. DE-AC02-05CH11231.

  \bibliographystyle{elsarticle-num-names} 
  \bibliography{bibfileTemplate}

\end{document}